\documentclass[onecolumn,12pt,draft]{ieeetran}
\usepackage{amsmath}
\usepackage[final]{graphicx}
\usepackage{amsmath,epsfig,cite}
\usepackage{graphicx}
\usepackage{amssymb}
\usepackage{mathrsfs}
\usepackage[dvips,dvipdfm]{geometry}
\usepackage{setspace}
\newtheorem{thm}{Theorem}

\newcommand{\pr}{\ensuremath{\text{Pr}}}

\newcommand{\dr}{d}
\newcommand{\du}{\partial}
\newcommand{\berx}{\ensuremath{{\text{P}_e(\rho x)}}}
\newcommand{\avber}{\ensuremath{\overline{\text{P}}_e(\rho, N)}}

\newcommand{\eber}{\ensuremath{\text{E}_{\mathcal{N}} \left[\overline{\text{P}}_e(\rho, \mathcal{N}) \right]}}
\newcommand{\beren}{\ensuremath{\overline{\text{P}}_e(\rho, \lambda)}}
\newcommand{\cdfrng}{\ensuremath{F_{\gamma_n}(x)}}
\newcommand{\cdfrn}{\ensuremath{F_{\gamma}(x)}}
\newcommand{\cdfstn}{\ensuremath{F_{\gamma^{\ast}}(x|\mathcal{N} = N)}}
\newcommand{\pdfrn}{\ensuremath{f_{\gamma}(x)}}
\newcommand{\avcap}{\ensuremath{\overline{C}(\rho, N)}}

\newcommand{\ecap}{\ensuremath{\text{E}_{\mathcal{N}} \left[\overline{C}(\rho, \mathcal{N})\right]}}
\newcommand{\capen}{\ensuremath{\overline{C}(\rho,\lambda)}}
\newcommand{\E}{\ensuremath{\text{E}}}

\newcommand{\ld}{\ensuremath{\lambda}}
\newcommand{\uld}{\ensuremath{\underline{\lambda}}}
\newcommand{\al}{\ensuremath{\alpha}}

\newcommand{\cdfinv}{\ensuremath{F_{\gamma}^{-1}(e^{-u})}}
\newcommand{\eu}{\ensuremath{e^{-u}}}

\begin{document}

\title{Multi-User Diversity with Random Number of Users}
\author{Adarsh B. Narasimhamurthy, Cihan Tepedelenlio\u{g}lu, \emph{Member, IEEE}, Yuan Zhang
\footnote{\scriptsize{A. B. Narasimhamurthy was with the School of
Electrical, Computer, and Energy Engineering, Arizona State
University, Tempe, AZ 85287, USA. He is now with The MathWorks, Inc.
(Email: adarsh.murthy@asu.edu). C. Tepedelenlio\u{g}lu and Y. Zhang
are with the School of Electrical, Computer, and Energy Engineering,
Arizona State University, Tempe, AZ 85287, USA. (Email:
cihan@asu.edu, yzhang93@asu.edu).}}}
\date{}
\maketitle

\vspace{-0.6in}

\begin{abstract}
Multi-user diversity is considered when the number of users in the
system is random.  The complete monotonicity of the error rate as a
function of the (deterministic) number of users is established and
it is proved that randomization of the number of users always leads
to deterioration of average system performance at any average SNR.
Further, using stochastic ordering theory, a framework for
comparison of system performance for different user distributions is
provided. For Poisson distributed users, the difference in error
rate of the random and deterministic number of users cases is shown
to asymptotically approach zero as the average number of users goes
to infinity for any fixed average SNR. In contrast, for a finite
average number of users and high SNR, it is found that randomization
of the number of users deteriorates performance significantly, and
the diversity order under fading is dominated by the smallest
possible number of users. For Poisson distributed users
communicating over Rayleigh faded channels, further closed-form
results are provided for average error rate, and the asymptotic
scaling law for ergodic capacity is also provided. Simulation
results are provided to corroborate our analytical findings.
\end{abstract}



\section{Introduction} \label{sec:intro}

Point to point diversity combining schemes aim to mitigate the
effects of fading in a wireless channel. In contrast, for multi-user
systems another form of diversity termed \emph{multi-user diversity}
(MUD) is available, which thrives on the randomness of the user
fading channels \cite{tse_viswanath}. The key idea is to provide
channel access to the user with the best channel at any instant of
time. This has been shown to be optimal for both uplink
\cite{knopp_humblet} and downlink \cite{tse_multiuser_div}
scenarios.

In the literature, MUD has been studied for the case of
deterministic number of users only. Since the number of users is
randomly varying in practice, it is of interest to consider MUD for
this case as well. For example, cell phone users have longer voice
calls while channel access for data communication is very short
\cite{cdma2000}. The probability of a cell phone user requesting
data communication is very low, and bursty data requests such as
stocks, weather and email lead to very short channel access times.
This suggests that the number of users actively contending for
channel access across time is random. Additionally, schemes in which
a user is allowed to feedback its channel estimate to request
channel access, when it is larger than a predefined threshold
\cite{Tang_heath_cho_yun,park_love, tang_heath_dlink}, also lead to
a random number of users. Even in common scenarios where the
fluctuations in the number of users is slower than the rapidity of
channel fading, averaging error rates, or ergodic capacity, with
respect to the user distribution results in meaningful system-level
performance measures.

In this paper, we analyze the performance of MUD systems with random
number of users for the first time in the literature. In Section
\ref{sec: System_model} the instantaneous SNR distribution of the
best user chosen from a random set of users is derived for arbitrary
fading and user distributions, and the mathematical preliminaries
are presented. In Section \ref{sec: char_ber_cap}, the error rate
averaged across fading with deterministic number of users is shown
to be a completely monotonic function of the number of users $N$.
Further, we also prove that the ergodic capacity of a MUD system
with a deterministic number of users has a completely monotone
derivative with respect to the number of users. These structural
results of performance for a deterministic number of users are then
used to prove facts about the random number of users case. The first
of these is that randomization of the number of users results in the
deterioration of average performance measured in terms of either
error rate or ergodic capacity, by using Jensen's inequality. In
Section \ref{sec: LT_ordering} we introduce a framework in which
different user distributions can be compared through the so called
Laplace transform partial ordering of the number of users which is a
particular stochastic order \cite{shaked_stochastic_1994}. In
Section \ref{sec: high_snr_anal} we derive the diversity order of a
MUD system with random number of users and show that it is
determined by the minimum possible number of users. In Section
\ref{sec: outage_poisson_general}, expressions for outage for any
fading distribution with Poisson user distribution are derived. For
when the user distribution is Poisson distributed, Jensen's
inequality for error rate is proved to be asymptotically tight in
the average number of users in Section \ref{sec: jensen_ber_tight}.
For the special case when the number of users is Poisson distributed
and when the user channel is Rayleigh faded, a closed-form
expression for the error rate is derived in Section \ref{sec:
poisson_exp}. The scaling of ergodic capacity with the average
number of users is also provided for this case in Section \ref{sec:
asymp_scaling_of_cap}. Poisson user distribution without allowing
the number of users to be zero is considered in Section
\ref{zero_trun}. Section \ref{sec: simulations} corroborates our
analytical results with simulations and Section \ref{sec:
conclusions} concludes the paper.

Here are some remarks on notations used in this work. Asymptotic
equivalence $\tau(x) \sim g(x)$ as $x \rightarrow a$ means that
$\lim_{x \rightarrow a} \tau(x)/g(x) = 1$, and $\tau(x) = O(g(x))$
as $x \rightarrow a$ means that $\limsup_{x \rightarrow a}
|\tau(x)/g(x)|< \infty$. In this paper, we consider $a=0$ or
$a=\infty$. ${\text{Pr}}[\cdot]$ is the probability of an event,
${\rm{E}}_{\mathcal{X}}[f(\mathcal{X})]$ is the expectation of the
function $f(\cdot)$ over the distribution of the random variable
$\mathcal{X}$, and $\log(\cdot)$ is logarithm to base $e$.

\section{System Model and Mathematical Preliminaries}\label{sec: System_model}

We consider an uplink MUD system with one base station (BS) and
multiple users. Without loss of generality, both the BS and the
users are assumed to have a single antenna. The received signal at
the BS from the $n^{th}$ user can be expressed as,
\begin{equation} \label{eqn: rcvd_sig}
y_n = \sqrt{\rho}h_n x_n+w_n, \hspace{0.4 in} n = 0, 1,\ldots,
\mathcal{N},
\end{equation}
where the number of users $\mathcal{N}$ is assumed to be a random
variable with a discrete non-negative integer distribution. In
Section \ref{zero_trun} we address the implications on the
performance metrics of allowing the probability of $\mathcal{N}=0$
to be positive, leading to possibly no users and no transmission.
When addressing the deterministic number of users case, we will set
$\mathcal{N} = N$, where $N$ is a realization of the random variable
$\mathcal{N}$. A homogeneous MUD system is assumed where the average
received power at the BS, $\rho$, is identical across all users. The
symbol $h_n$ denotes the channel coefficient, $x_n$ the transmitted
symbol, and $w_n$ the additive white Gaussian noise (AWGN)
corresponding to the $n^{th}$ user. The channel is assumed to
satisfy ${\rm{E}}[|h_n|^2] = 1$ for all $n$ and to be independent
and identically distributed (i.i.d.) across all users. The
transmitted symbols satisfy ${\rm{E}}[|x_n|^2] = 1$.

The channel gain of the $n^{th}$ user at the BS, prior to selection,
can be expressed as $\gamma_n = |h_n|^2$, and the selected user has
a channel gain denoted by ${\gamma}^{\ast} = |h_{\ast}|^2$, where
$|h_{\ast}|^2 = \max_{n}\{|h_{n}|^2\}$. Note that $\gamma^{\ast}$ is
a random variable that depends on the random variables
$\mathcal{N}$, and $|h_n|^2$, $n=0, 1, \ldots, \mathcal{N}$.

Define $\cdfrng$ as the cumulative distribution function (CDF) of
the channel gain of the $n^{th}$ user $\gamma_{n}$. Since the fading
channels across all users are assumed to be i.i.d., we drop the
index $n$ and define $\cdfrn := \cdfrng$. Recalling that the total
number of users $\mathcal{N}$ is a random variable, the CDF of the
channel gain of the selected user, conditioned on $\mathcal{N} = N$,
can be written as:
\begin{equation}\label{eqn: cdf_condn}
\cdfstn = F_{\gamma}^N(x),
\end{equation}
where the $N^{th}$ power is obtained due to the i.i.d. assumption of
the $N$ user channels. The CDF of the channel gain of the best user
selected from a random set of users can be obtained by averaging
\eqref{eqn: cdf_condn} with respect to the distribution of
$\mathcal{N}$:
\begin{equation} \label{eqn: cdf_gen}
F_{\gamma^{\ast}}(x)= {\rm{E}}_{\mathcal{N}}\left[
F_{\gamma}^{\mathcal{N}}(x) \right]= \sum_{k = 0}^{\infty}
{\pr}\left[\mathcal{N}= k \right] F_{\gamma}^k(x)=
U_{\mathcal{N}}(\cdfrn)
\end{equation}
where $U_{\mathcal{N}}(t) = \sum_{k=0}^{\infty}
{\pr}\left[\mathcal{N} = k \right]t^k$, $0\leq t\leq 1$, is the
probability generating function (PGF) of random variable
$\mathcal{N}$. From \eqref{eqn: cdf_gen} it can be seen that for any
fading channel distribution and any non-negative integer
distribution on the number of users, the CDF of the best user's
channel gain at the BS can be easily obtained.

We now survey some mathematical preliminaries that will be useful
throughout. A function $\tau(x):\mathbb{R}^+ \rightarrow \mathbb{R}$
is {\it completely monotonic} ($c.m.$) if its derivatives alternate
in sign \cite{shaked_stochastic_1994}, i.e.,
\begin{equation}\label{eqn: math_prelim_one}
(-1)^k \frac{d^k \tau(x)}{\dr x^k} \geq 0, \hspace{0.2in} \forall x,
\hspace{0.2in} k = 0, 1, 2, \ldots,
\end{equation}
where $d^0 \tau (x)/dx^0= \tau (x)$ by definition. Due to a
celebrated theorem by Bernstein \cite{shaked_stochastic_1994}, an
equivalent definition for $c.m.$ is that it is a positive mixture of
decaying exponentials. In other words, we have the Bernstein's
representation $\tau(x) = \int_0^{\infty} e^{-sx} d\psi(s)$ for some
nondecreasing function $\psi(s)$. In this paper, we are sometimes
interested in $c.m.$ functions on integers, which are nothing but
sequences obtained by sampling $c.m.$ functions as defined by
\eqref{eqn: math_prelim_one}. We are also interested in functions
whose first-order derivatives satisfy \eqref{eqn: math_prelim_one},
which are said to have a completely monotone derivative ($c.m.d.$).
Even when the variable $x$ is naturally an integer (such as the
number of users), we will sometimes treat it as a real number, since
we will be primarily interested in the asymptotic properties of
$\tau(x)$.

A function $\psi(s)$ is {\it regularly varying} with exponent $\mu
\neq 0$ at $s=\infty$ if it can be expressed as $\psi(s)= s^{\mu}
l(s)$ where $l(s)$ is slowly varying and by definition satisfies
$\lim_{s \rightarrow \infty} l(\kappa s)/l(s)= 1$ for $\kappa>0$.
Regular (slow) variation of $\psi(s)$ at $s=0$ is equivalent to
regular (slow) variation of $\psi(1/s)$ at $\infty$. Intuitively,
regular variation captures polynomial-like behavior near the origin
or at infinity. The Tauberian theorem for Laplace transforms, whose
proof can be found in \cite{feller_introduction_2009}, applies to
$c.m.$ functions and states that $\tau(x)$ is regularly varying at
$x=\infty$ if and only if $\psi(s)$ is regularly varying at $s=0$:
\begin{thm} \label{tbr_th}
If a nondecreasing function $\psi(s) \geq 0$ defined on $s \geq 0$
has Laplace transform $\tau(x) = \int_0^{\infty} e^{-sx} d\psi(s)$
for $x \geq 0$, and $l(s)$ is slowly varying at $s=0$ (or
$s=\infty$), the relations $\psi(s) \sim s^{\mu} l(s)$ as $s
\rightarrow 0$ (or $s \rightarrow \infty$) and $\tau(x) \sim
\Gamma(\mu+1) x^{-\mu} l(x^{-1})$ as $x \rightarrow \infty$ (or $x
\rightarrow 0$) imply each other, where $\mu \in \mathbb{R}$.
\end{thm}

One useful property given in \cite[p.27]{bingham_regular_1989} is
that
\begin{equation} \label{sv_out_int}
\int_0^t s^{\mu} l(s) ds \sim l(t) \int_0^t s^{\mu} ds=
\frac{1}{\mu+1} t^{\mu+1} l(t)
\end{equation}
as $t \rightarrow 0$, with $\mu>-1$ for $l(s)$ slowly varying at
$s=0$.

In this paper, we are interested in studying average error rates,
and capacities averaged across both the channel distribution, and
the number of users. The expression $\avber$ represents the error
rate of a MUD system with a deterministic number of users $N$, that
is averaged with respect to the distribution of the fading channel.
The expression $\eber$ represents the average error rate of a MUD
system with a random number of users, which is averaged with respect
to the distribution of the number of users {\it and} the fading
channels.

\section{Properties of the Average Error Rate and Ergodic Capacity}\label{sec: char_ber_cap}

\subsection{Average Error Rate} \label{sec: ber_cmf}

In this section, we first prove that the average error rate of a MUD
system, with a deterministic number of users $N$, is a $c.m.$
function of $N$, under general conditions. This will be used to
infer about the behavior of the average error rate when a random
number of users is considered, in Section \ref{sec: LT_ordering}.

The error rate of a MUD system with a deterministic number of users
$N$ and average SNR $\rho$ is given by,
\begin{equation} \label{eqn: avg_BER1}
\avber = \int_0^{\infty} \berx \dr F_{\gamma}^N (x)
\end{equation}
where $\berx$ is the instantaneous error rate over an AWGN channel
for an instantaneous SNR $\rho x$ of the best user. Often, the
instantaneous error rate is assumed to have the form $\berx = \al
e^{-\eta \rho x}$ or $\berx = \al Q(\sqrt{\eta \rho x})$, where
$\alpha$ and $\eta$ can be chosen to capture different modulations
\cite{goldsmith}. To represent \eqref{eqn: avg_BER1} in terms of the
CDF $\cdfrn$, rather than the probability density function (PDF), we
left it as a Stieltjes integral \cite{abramowitz_handbook_1964} even
though it can also be expressed in terms of the PDF $\pdfrn$ using
$\dr F_{\gamma}^N (x) = N F_{\gamma}^{N-1} (x) \pdfrn$. In what
follows, we will study the sequence $\avber$ as a function of the
integer variable $N$. Since we are ultimately interested in the
asymptotic behavior of this sequence, we will also consider
\eqref{eqn: avg_BER1} with $N$ being a real number.

We begin by proving that $\avber$ is a $c.m.$ function of $N$ not
just for $\berx$ in the forms of exponential function and Q
function, but for any instantaneous error rate function. In other
words, we only assume $\berx$ is decreasing in $x$ for any $\rho>0$.
Defining $B(x)= -\dr \text{P}_e(x)/\dr x$, after integrating
\eqref{eqn: avg_BER1} by parts, the $k^{th}$ derivative of $\avber$
can be written as,
\begin{equation}\label{eqn: ber_dr1}
\frac{\du^{k}\avber}{\du N^{k}} = \rho \int_0^{\infty} B(\rho x)
F_{\gamma}^N(x) \left[\log\left(\cdfrn\right)\right]^k \dr x.
\end{equation}
Since $\berx$ is decreasing, and $\log\left(\cdfrn \right) \leq 0$
we see that \eqref{eqn: ber_dr1} satisfies the definition in
\eqref{eqn: math_prelim_one}. In particular, $\avber$ being a $c.m.$
function means that \eqref{eqn: ber_dr1} is negative for $k=1$ and
positive for $k=2$, and consequently $\avber$ is a convex decreasing
function of $N$. For when the number of users in the system is
random, by applying Jensen's inequality for convex functions, we
have,
\begin{equation}
\eber \geq \beren,
\end{equation}
where $\ld := \rm{E}[\mathcal{N}]$. Therefore, randomization of the
number of users always deteriorates the average error rate
performance of a MUD system.

To establish the complete monotonicity of $\avber$ as a function of
$N$, we only used the fact that the instantaneous error rate $\berx$
in \eqref{eqn: avg_BER1} is a decreasing function of $x$ for
$\rho>0$, which always holds. This $c.m.$ property will be used to
stochastically order user distributions in Section \ref{sec:
LT_ordering}.

\subsection{Ergodic Capacity}\label{sec: cap_cmd}

The ergodic capacity for the deterministic number of users system
can be expressed as,
\begin{align}
\avcap = \int_0^{\infty} \log\left(1+\rho x \right) \dr F_{\gamma}^N
(x)= \rho \int_0^{\infty} \frac{1-F_{\gamma}^N (x)}{1+\rho x} \dr x
\label{eqn: cap_av1}.
\end{align}
where we use integration by parts, and assume that $F_{\gamma} (x)$
satisfies $\lim_{x \rightarrow \infty} \log(1+\rho x)(1-F_{\gamma}^N
(x)) = 0$, for all $N \geq 0$.

It can be seen that $\avcap$ has a completely monotonic derivative
since,
\begin{equation}\label{eqn: avcap_cmf}
\frac{\du^{k+1} \avcap}{\du N^{k+1}} = -\rho \int_0^{\infty}
\frac{F_{\gamma}^N (x) \left[\log\left(\cdfrn
\right)\right]^{k+1}}{1+\rho x} \dr x.
\end{equation}
alternates in sign as $k$ is incremented. This establishes that
$\avcap$ has a completely monotonic derivative, provided that the
fading distribution satisfies the mild assumption $\lim_{x
\rightarrow 0} \log(1+\rho x)(1-F_{\gamma}^N (x)) = 0$ for all $N
\geq 0$, as assumed after \eqref{eqn: cap_av1}. This assumption
holds for all distributions with exponential or power law tails,
which is the case for all fading distributions in wireless
communications. Using \eqref{eqn: avcap_cmf} with $k=0,1$, it is
seen that $\avcap$ is concave increasing function of $N$. Applying
Jensen's inequality for concave functions, we have
\begin{equation} \ecap \leq \capen.
\end{equation}
Therefore, similar to the error rate metric, randomization of $N$
will always hurt the average ergodic capacity of a MUD system. The
$c.m.d.$ property of $\avcap$ will be used in the following section
discussing the stochastic Laplace transform ordering of user
distributions.

\section{Laplace Transform Ordering of User Distributions}\label{sec: LT_ordering}

In this section we introduce Laplace transform (LT) ordering, a tool
to compare the effect that different user distributions has on the
error rate, and ergodic capacity averaged across user and channel
distributions. Stochastic ordering of random variables, of which LT
ordering is a special case, is a branch of probability theory and
statistics which deals with binary relations between random
variables \cite{shaked_stochastic_1994, mueller_comparison_2002}.

Let $\mathcal{X}$ and $\mathcal{Y}$ be non-negative random
variables. $\mathcal{X}$ is said to be less than $\mathcal{Y}$ in
the LT order (written $\mathcal{X} \leq_{Lt} \mathcal{Y}$), if
$\E\left[e^{-s\mathcal{X}} \right] \geq \E\left[ e^{-s\mathcal{Y}}
\right]$ for all $s>0$. An important theorem found in
\cite{shaked_stochastic_1994}, and \cite{mueller_comparison_2002} is
given next:
\begin{thm}\label{thm: one}
Let $\mathcal{X}$ and $\mathcal{Y}$ be two random variables. If
$\mathcal{X} \leq_{Lt} \mathcal{Y}$, then,
$\E\left[\psi(\mathcal{X}) \right] \geq \E\left[ \psi(\mathcal{Y})
\right]$ for all $c.m.$ functions $\psi(\cdot)$, provided the
expectation exists. Moreover, the reverse inequality
$\E[\psi(\mathcal{X})] \leq \E[\psi(\mathcal{Y})]$ holds for all
$\psi(\cdot)$ with a completely monotone derivative, provided the
expectation exists.
\end{thm}

It follows that if two user distributions satisfy $\mathcal{N}_1
\leq_{Lt} \mathcal{N}_2$, then for all average SNR $\rho$
\begin{align}\label{eqn: jensens_ineq}
\E_{\mathcal{N}_1}\left[ \overline{\text{P}}_e(\rho, \mathcal{N}_1) \right]
&\geq \E_{\mathcal{N}_2} \left[\overline{\text{P}}_e(\rho, \mathcal{N}_2)\right], \nonumber \\
\E_{\mathcal{N}_1}\left[ \overline{C}(\rho, \mathcal{N}_1) \right]
&\leq \E_{\mathcal{N}_2} \left[  \overline{C}(\rho,
\mathcal{N}_2)\right].
\end{align}
To rephrase \eqref{eqn: jensens_ineq}, if the number of users
$\mathcal{N}$ is from a distribution that is dominated by another
distribution in the Laplace transform sense, then both the average
error rate and capacity are respectively ordered at all average SNR
$\rho$.

The LT ordering of discrete random variables can also be expressed
in terms of the ordering of their PGFs. By defining $t:=e^{-s}$, one
can rewrite $\E\left[e^{-s\mathcal{X}}\right] \geq
\E\left[e^{-s\mathcal{Y}}\right]$ for $s \geq 0$ as
$\E\left[t^{\mathcal{X}}\right] \geq \E\left[t^{\mathcal{Y}}\right]$
for $0\leq t \leq 1$, which is the same as $U_{\mathcal{X}}(t) \geq
U_{\mathcal{Y}}(t)$, $0 \leq t \leq 1$, where we recall that
$U_{\mathcal{X}}(t) = \E[t^\mathcal{X}]$ represents the PGF of the
discrete random variable $\mathcal{X}$.

To provide examples of random variables that are LT ordered,
consider Poisson random variables $\mathcal{X}$ and $\mathcal{Y}$
with means $\ld$ and $\mu$ respectively, such that $\ld \leq \mu$.
It is straightforward to show that for this case $e^{\ld (t-1)}=
\E\left[t^{\mathcal{X}} \right] \geq \E\left[t^{\mathcal{Y}}
\right]= e^{\mu (t-1)}$, for $0 \leq t \leq 1$, implying that
$\mathcal{X} \leq_{Lt} \mathcal{Y}$. If $\mathcal{X}$ and
$\mathcal{Y}$ are geometric distributed with probability of success
on each trial $p_1$ and $p_2$ respectively, such that $p_1 \leq
p_2$, then $\mathcal{Y} \leq_{Lt} \mathcal{X}$ since $p_2/(1-(1-p_2)
t)= \E\left[t^{\mathcal{Y}}\right] \leq
\E\left[t^{\mathcal{X}}\right] =p_1/(1-(1-p_1) t)$ for $0 \leq t
\leq 1$. Similarly, for $\mathcal{X}$ being Poisson distributed with
parameter $\ld$ and $\mathcal{Y}$ being geometric distributed with
parameter $p=1/(1+\ld)$, so that ${\rm{E}}[\mathcal{X}] =
{\rm{E}}[\mathcal{Y}] = \ld$, it can be once again shown that
$\E\left[t^{\mathcal{Y}}\right] \leq \E\left[t^{\mathcal{X}}\right]$
for $0 \leq t \leq 1$, establishing $\mathcal{Y} \leq_{Lt}
\mathcal{X}$. From this latter result, one can conclude that Poisson
offers a better user distribution than geometric distribution for a
fixed average number of users at all average SNR $\rho$, from both
error rate and capacity points of view.

\section{High SNR Analysis and Diversity Order}\label{sec: high_snr_anal}

In this section we analyze the average error rate at high SNR under
general assumptions on the user distribution and fading channel
distribution. In the following, we will assume that the fading
distribution $F_{\gamma}(x)$ is regularly varying with exponent $d$
at $x=0$ (typically $F_{\gamma}(x)= O(x^d)$ as $x \rightarrow 0$),
which is true for many fading distributions including Rayleigh
($d=1$), Nakagami-$m$ ($d=m$) and Ricean ($d=1$) \cite{wang03}.

\begin{thm} \label{thm_ser_asympt}
Let $F_{\gamma}(x)$ be regularly varying at $x=0$ with exponent
$d>0$, and $\mathcal{N} \in \{ k_0, k_0+1, ... \}$ represent the
range of the number of users random variable. Then the high-SNR
asymptotic average error rate is given by
\begin{equation} \label{ser_asympt}
\E_{\mathcal{N}}\left[ \overline{\text{P}}_e(\rho, \mathcal{N})
\right] \sim \pr\left[\mathcal{N} = k_0 \right] C_1
F_{\gamma}^{k_0}(C_2 \rho^{-1})
\end{equation}
as $\rho \rightarrow \infty$, where $C_1$ and $C_2$ are constants
given by $C_1=\alpha \Gamma(k_0 d+ 1)$, $C_2=\eta^{-1}$ when $\berx
= \al e^{-\eta \rho x}$ and $C_1=\alpha \Gamma(k_0 d+ 1/2)/(2
\sqrt{\pi})$, $C_2=2\eta^{-1}$ when $\berx = \al Q(\sqrt{\eta \rho
x})$.
\end{thm}

\begin{proof}
See \ref{thm_ser_asympt_proof}.
\end{proof}

From \eqref{ser_asympt} it is straightforward to show that the
diversity order of the MUD system with random number of users is
given by $k_0 d$. This follows from \eqref{ser_asympt} and the
regular variation assumption on $F_{\gamma}(x)$, so that
$F_{\gamma}^{k_0}(C_2 \rho^{-1})= O(\rho^{-k_0 d})$ as $\rho
\rightarrow \infty$.

\section{Poisson Distributed $\mathcal{N}$} \label{sec: poisson_dsbn_N}

Consider a MUD system which contains a large number of users.
Suppose each user is active with a small probability. In such a
system, as the number of users increases, the user distribution will
approximate the Poisson. In this section we analyze the system when
$\mathcal{N}$ is Poisson distributed with parameter $\ld$.

\subsection{Outage Probability and Its Asymptotic Behavior for Large $\ld$} \label{sec: outage_poisson_general}

When $\mathcal{N}$ is Poisson distributed with parameter $\ld$, the
probability of outage with a threshold $x$ can be expressed as,
\begin{equation} \label{eqn: outage_poiss_gen}
F_{\gamma^{\ast}}(x)=\pr\left[\gamma^{\ast} \leq x\right] =
\sum_{k=0}^{\infty} e^{-\ld} \frac{\ld^k}{k!} F_{\gamma}^k(x) =
e^{-\ld(1-F_{\gamma}(x))} I[x \geq 0]
\end{equation}
where $I[\cdot]$ is the indicator function. Equation \eqref{eqn:
outage_poiss_gen} implies that as the average number of users
increases, the outage probability decreases for any distribution on
$\gamma_n$, the channel gain of the user fading channel.

In what follows, we show that for large $\ld$ the outage behavior is
dependent on $\cdfrn$ only through its tail behavior. In fact, it is
possible to show that there exist normalizing and shift functions
$a(\ld)$ and $b(\ld)$ such that the probability $\pr\left[
(\gamma^{\ast}-b(\ld))/a(\ld) \leq x\right]$ for large $\ld$ is
\begin{equation} \label{gumbel_param1}
\lim_{\ld \rightarrow \infty} F_{\gamma^{\ast}}(a(\ld)x+ b(\ld))=
\exp(-e^{-x}), \hspace{1cm} -\infty<x<\infty
\end{equation}
which is known as the Gumbel distribution \cite{david}. Using
\eqref{eqn: outage_poiss_gen}, sufficient and necessary conditions
for \eqref{gumbel_param1} are clearly that $\lim_{\ld \rightarrow
\infty} \ld (1- F_{\gamma}(a(\ld)x+ b(\ld)))= e^{-x}$. In
\cite[p.300]{david} it is shown that $a(\ld)=[\ld f_{\gamma}
(b(\ld))]^{-1}$, $b(\ld)= F_{\gamma}^{-1}(1-1/\ld)$, satisfy this
condition for many distributions including Rayleigh, Nakagami-$m$
and Ricean. Also, it can be seen that the asymptotic CDF of the
Poisson number of users case for large $\ld$ has the same form
(Gumbel distribution) as the asymptotic CDF of the deterministic
number of users case for large $N$.

\subsection{Average Error Rate} \label{sec: jensen_ber_tight}

In our outage analysis for the Poisson number of users case, we
showed that the outage probability in \eqref{eqn: outage_poiss_gen}
for large $\ld$ approaches the Gumbel distribution, which is also
the asymptotic distribution obtained for a deterministic number of
users case. Therefore, even though we saw that randomization always
deteriorates performance, for large average number of users it
should approximately yield the same performance as the deterministic
case. This amounts to the tightness of Jensen's inequality for the
Poisson users case.

We now provide sufficient conditions for Jensen's inequality
involving $\avber$ in \eqref{eqn: jensens_ineq} to be asymptotically
tight in $\ld$. Recall that $\avber$ is the error rate averaged over
the channel distribution for deterministic number of users $N$. To
this end, we use the results in \cite[Theorem
2.2]{Downey93anabelian} which were derived in a networking context
for arbitrary {\it c.m.} functions.

\begin{thm}\label{thm: four}
Let $\avber$ be $c.m.$ and regularly varying at $N=\infty$ and
consider the error rate averaged across the channel and the users
$\eber$, where $\mathcal{N}$ is a Poisson distributed random
variable with mean $\ld$. Then,
\begin{equation}\label{eqn: jsens_tight}
\eber = \beren+O\left(\overline{\text{P}}_e(\rho, \ld)/\ld \right)
\end{equation}
as $\ld \rightarrow \infty$.
\end{thm}

Equation \eqref{eqn: jsens_tight} shows that as $\ld \rightarrow
\infty$, the difference between the error rate averaged across the
user distribution and the error rate evaluated at the average number
of users vanishes as $\ld$ tends to $\infty$. This implies that for
sufficiently large $\ld$ the performance of the MUD systems with
random number of users will be almost equal to the performance of
the MUD systems with a deterministic number of users with the number
of users equal to $\ld$.

To apply Theorem \ref{thm: four} we require $\avber$ to be $c.m.$
and regularly varying. We have already shown that $\avber$ is always
completely monotonic in $N$. Next, we provide the conditions under
which $\avber$ is a regularly varying function of $N$. Consider
\begin{equation}
\avber = \rho \int_0^{\infty} B(\rho x) e^{N \log(\cdfrn)} \dr x
\end{equation}
where $B(\cdot)$ is defined as $B(x)= -\dr \text{P}_e(x)/\dr x$.
Now, setting $u := -\log(\cdfrn)$, and integrating by substitution
we have,
\begin{equation} \label{eqn: downey_res3}
\avber = \rho \int_0^{\infty} \frac{B(\rho \cdfinv) e^{-u} e^{-u N}
\dr u}{f_{\gamma}(\cdfinv)},
\end{equation}
where $F_{\gamma}^{-1}(x)$ is the inverse CDF and $\pdfrn$ is the
PDF of $\gamma_n$. We now establish the sufficient conditions for
$\avber$ to be a regularly varying function of $N$:

\begin{thm}\label{thm: five}
If $\avber$ is $c.m.$ in $N$, a sufficient condition for it to be
regularly varying at $N=\infty$ is that, $t(u):=\rho (B(\rho
\cdfinv)e^{-u})/(f_{\gamma}\left(\cdfinv\right))$ is regularly
varying at $u=0$.
\end{thm}

\begin{proof}
By comparing the representation of $\avber$ in \eqref{eqn:
downey_res3} with the Bernstein's representation of $c.m.$ functions
discussed after \eqref{eqn: math_prelim_one}, it can be seen that
\eqref{eqn: downey_res3} can be represented as the Laplace transform
of $t(u)$. Using Theorem \ref{tbr_th}, the proof follows.
\end{proof}

Theorem \ref{thm: five} shows that for the conclusions of Theorem
\ref{thm: four} to hold (i.e., Jensen's inequality to be
asymptotically tight), the CDF of the single-user channel
$F_\gamma(x)$, and the error rate expression $\berx$ have to jointly
satisfy the regular variation condition given in Theorem \ref{thm:
five}. Next, we examine whether this condition holds for commonly
assumed instantaneous error rates $\berx$ with $\gamma_n$ being
exponentially distributed. For the case of $\berx= \alpha e^{-\eta
\rho x}$, we have $t(u)= \alpha \rho (1-\eu)^{\eta \rho-1} \eu$,
which satisfies $\lim_{u \rightarrow 0} t(\kappa u)/t(u) =
\kappa^{\eta \rho-1}$, therefore proving the regular variation of
$t(u)$ at $0$. By using Theorem \ref{tbr_th} this in turn proves
regular variation of $\avber$ at $N=\infty$. Therefore $\avber$ is
both a $c.m.$ and a regularly varying function of $N$ for this case.
Consequently, when $\berx= \alpha e^{-\eta \rho x}$ and the fading
is Rayleigh (i.e. channel gain is exponential), the difference in
error rate performance of a MUD system with a random number of users
averaged over the number of users distribution and of a
deterministic number users approaches zero for sufficiently large
$\ld$, as in Theorem \ref{thm: four}.

Consider now $\berx = \alpha Q(\sqrt{\eta \rho x})$, with $\gamma_n$
being exponentially distributed. The error rate can be expressed as,
\begin{equation} \label{eqn: ex_exp1}
\avber = \alpha \int_0^{\infty} Q\left(\sqrt{\eta \rho x}\right)\dr
F_{\gamma}^N(x) = \frac{\alpha \sqrt{\eta \rho}}{2 \sqrt{2\pi}}
\int_0^{\infty} \frac{e^{N\log\left(1-e^{-x}\right)} e^{-\eta \rho
x/2}}{\sqrt{x}} \dr x,
\end{equation}
where the second equality is obtained by integration by parts. Once
again, by setting $u = -\log(1-e^{-x})$ we can rewrite \eqref{eqn:
ex_exp1} as,
\begin{equation}
\frac{\alpha \sqrt{\eta \rho}} {2\sqrt{2\pi}} \int_0^{\infty}
\exp\left(-Nu\right) (1-e^{-u})^{\eta \rho/2-1}
\frac{e^{-u}}{\sqrt{-\log(1-e^{-u})}} \dr u.
\end{equation}
Thus we have $t(u)=\alpha \sqrt{\eta \rho} (1-e^{-u})^{\eta
\rho/2-1}e^{-u}/(2\sqrt{-2\pi \log(1-e^{-u})})$ and it can be shown
that $\lim_{u \rightarrow 0} t(\kappa u)/t(u) = \kappa^{\eta
\rho/2-1}$, therefore once again proving that $\avber$ is both a
$c.m.$ and a regularly varying function of $N$. Having verified the
conditions of Theorem \ref{thm: five} for $\berx = \alpha
Q(\sqrt{\eta \rho x})$ with $\gamma_n$ being exponentially
distributed, we conclude the tightness of Jensen's inequality as
suggested by Theorem \ref{thm: four}.

\subsection{A Special Case: Poisson distributed $\mathcal{N}$ and Rayleigh Faded Channels}\label{sec: poisson_exp}

In this section, we consider the case when the number of users
$\mathcal{N}$ is Poisson distributed and the user channels are
Rayleigh faded. This practically relevant case will lead to closed
form expressions.

\subsubsection{Distribution of Channel Gain}\label{sec: poisson_exp_dsbn_of_snr}

For this case the CDF of the channel gain of the best user using
\eqref{eqn: outage_poiss_gen} is given by,
\begin{equation} \label{eqn: cdf_poisson_exp}
F_{\gamma^{\ast}}(x) = \exp\left(-\ld e^{-x}\right) I[x \geq 0].
\end{equation}
The channel gain of the best user in \eqref{eqn: cdf_poisson_exp} is
identical to a truncated Gumbel distribution, which was seen in its
untruncated form in \eqref{gumbel_param1}. Notice that for $x = 0$
\eqref{eqn: cdf_poisson_exp} yields $e^{-\ld}>0$ so
$F_{\gamma^{\ast}}(x)$ has a jump at $x=0$. The distribution in
\eqref{eqn: cdf_poisson_exp} is therefore of mixed type with a mass
of $e^{-\ld}$ at the origin and the rest of the distribution has the
form of a \emph{truncated Gumbel distribution}.

\subsubsection{Average Error Rate}

Assuming the error rate has the form, $\berx = \al e^{-\eta \rho x}$
as mentioned in Section \ref{sec: ber_cmf}, the average error rate
can be expressed as,
\begin{equation}\label{eqn: ber_poisson_exp}
\eber = \ld \al \int_0^{\infty} e^{-\eta \rho x} e^{-x} e^{-\ld
e^{-x}} \dr x+\al \int_0^{\infty}\delta(x)e^{-\ld}e^{-\eta \rho x}
\dr x.
\end{equation}
Setting $y = \ld e^{-x}$ and integrating by substitution,
\eqref{eqn: ber_poisson_exp} can be expressed as,
\begin{equation} \label{eqn: avg_ber_poiss_exp2}
\eber = {\al}\int_0^{\ld} \left(\frac{y}{\ld}\right)^{\eta \rho}
e^{-y} \dr y +\al e^{-\ld}={\al \ld^{-\eta \rho}} \gamma(\eta
\rho+1, \ld)+\al e^{-\ld},
\end{equation}
where $\gamma(s,x)$ is the lower incomplete gamma function
\cite{abramowitz_handbook_1964}. It can be easily shown that ${\al
\ld^{-\eta \rho}} \gamma(\eta \rho+1, \ld)+\al e^{-\ld} \sim {\al
\ld^{-\eta \rho}} \Gamma(\eta \rho+1)$ as $\ld \rightarrow \infty$,
indicating a power-law decay in the error rate as a function of the
average number of users.

\subsection{Asymptotic Scaling of Capacity with $\ld$} \label{sec: asymp_scaling_of_cap}

Next, we derive the asymptotic average capacity and the
corresponding scaling laws with respect to $\ld$.
\begin{thm} \label{cap_scaling}
For Poisson distributed $\mathcal{N}$ with mean $\ld$ and Rayleigh
faded channels, as $\ld \rightarrow \infty$, we have
\begin{equation} \label{cap1_asympt}
\ecap = \log\left(1+ \rho \log(\ld) \right) + O(1/\sqrt{\log(\ld)}).
\end{equation}
\end{thm}

\begin{proof}
See \ref{cap_scaling_proof}.
\end{proof}

For a MUD system with deterministic number of users $N$, it has been
shown in \cite{tse_viswanath} that the ergodic capacity grows as
$\log\log(N)$. From Theorem \ref{cap_scaling} it is seen that for a
MUD system with random number users, whose mean is $\ld$, the
ergodic capacity grows as $\log\log(\ld)$. This implies that when
average number of users $\ld$ is equal to $N$ of the deterministic
number of users case, the ergodic capacity for both cases grow at
the same rate.

\subsection{Zero Truncated Poisson User Distribution} \label{zero_trun}

The CDF expression in \eqref{eqn: cdf_poisson_exp} includes the case
when $N=0$, i.e., there are no users in the system. When there are
no users, no data will be transmitted. In view of this, it is
reasonable to drop the $N=0$ case and model the user distribution
with the \emph{zero-truncated Poisson distribution} which is given
by ${\pr} [\mathcal{N}=k| \mathcal{N}>0]= {\pr} [\mathcal{N}=k]/(1-
{\pr} [\mathcal{N}=0])$, for any positive integer $k$. For
zero-truncated Poisson distributed $\mathcal{N}$, $N \in \{1, 2,
\ldots\}$ and mean $\uld = \ld/(1-e^{-\ld})$ where $\ld$ is the mean
of the underlying Poisson random variable. The CDF of the channel
gain of the best user can be expressed as,
\begin{equation} \label{eqn: cdf_zero_truncated}
F_{\gamma^{\ast}}(x) = \frac{1}{1-e^{-\uld}}\sum_{k=1}^{\infty}
\left[\cdfrn \right]^k \uld^k \frac{e^{-\uld}}{k!}=
\frac{e^{\uld(1-e^{-x})}-1}{e^{\uld}-1}.
\end{equation}
For when the average number of users $\ld \rightarrow \infty$, it
can be seen that $\uld \sim \ld$. Further,
\begin{equation}
\lim_{\uld \rightarrow \infty} F_{\gamma^{\ast}}(x) = \lim_{\uld
\rightarrow \infty} \frac{e^{\uld(1-e^{-x})}-1}{e^{\uld}-1} =
e^{-\ld(e^{-x})},
\end{equation}
which is identical to the CDF for Poisson distributed user case in
\eqref{eqn: cdf_poisson_exp}. This implies that for a large average
number of users in the system, the outage, average error rate, and
ergodic capacity performance of zero-truncated Poisson distributed
user case will be identical to that of the Poisson distributed user
case.

\section{Simulations}\label{sec: simulations}

An uplink MUD system where both the BS and users having a single
antenna is considered. In this section, using Monte-Carlo
simulations, the error rate, ergodic capacity and outage capacity
are simulated to corroborate our analytical results. For all
simulations considered, the Rayleigh fading is assumed.

\begin{figure}[!htp]
\begin{center}
\includegraphics[height=9cm,keepaspectratio]{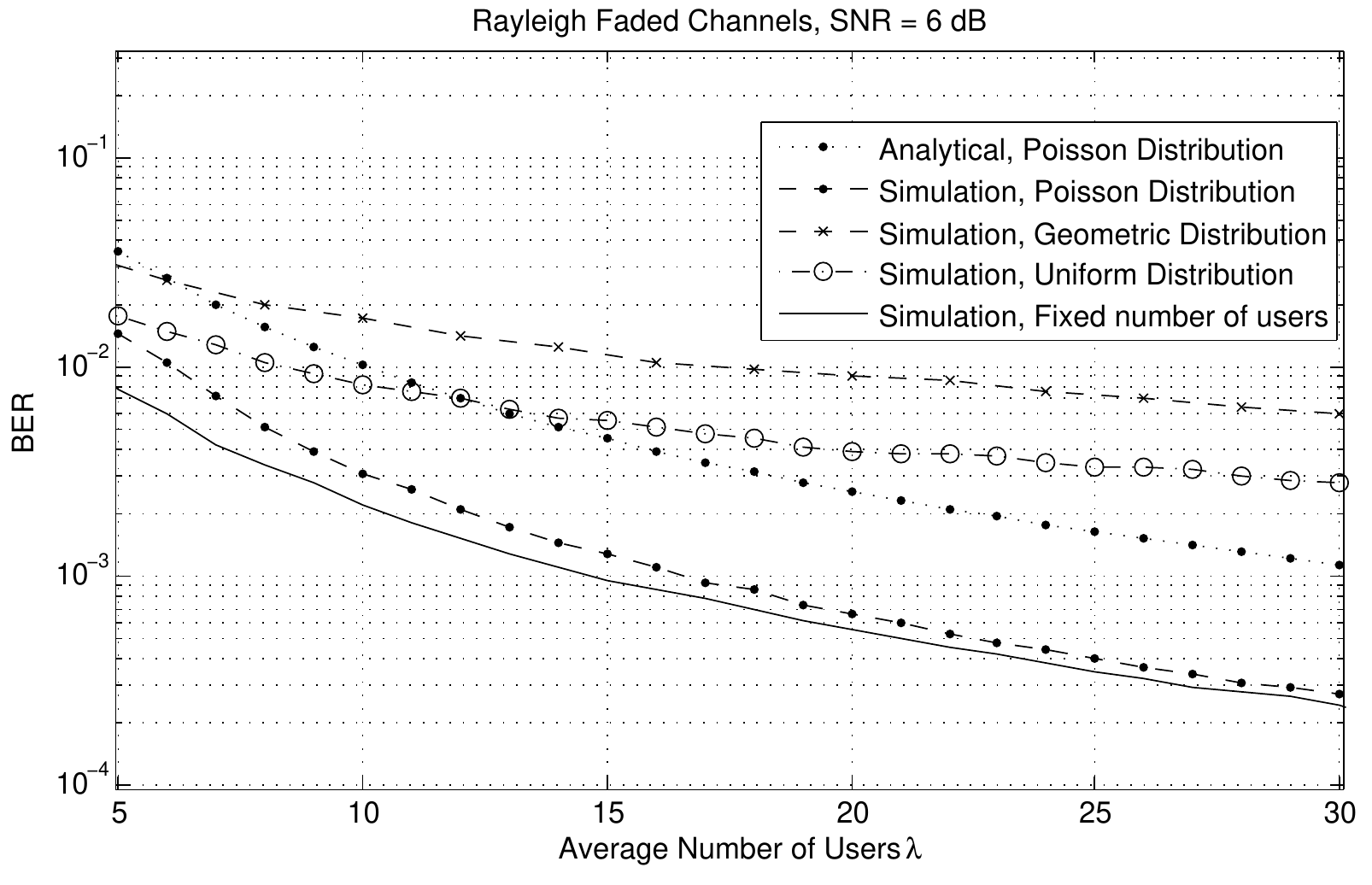}
\caption{Error rate vs. $\lambda$: Rayleigh Fading Channel, average
SNR = 6 dB} \label{fig: fig1}
\end{center}
\end{figure}

In Figure \ref{fig: fig1}, assuming $\pi/4$ QPSK modulation, the
average bit error rate with deterministic $N$ is compared with the
performance averaged across various user distributions. It is seen
that the deterministic number of users system performs better than
all the cases involving random number of users. The performance of
the Poisson distributed users case comes close to the deterministic
case as $\lambda$ increases, as predicted by Theorem \ref{thm:
four}.

\begin{figure}[!htp]
\begin{center}
\includegraphics[height=9cm,keepaspectratio]{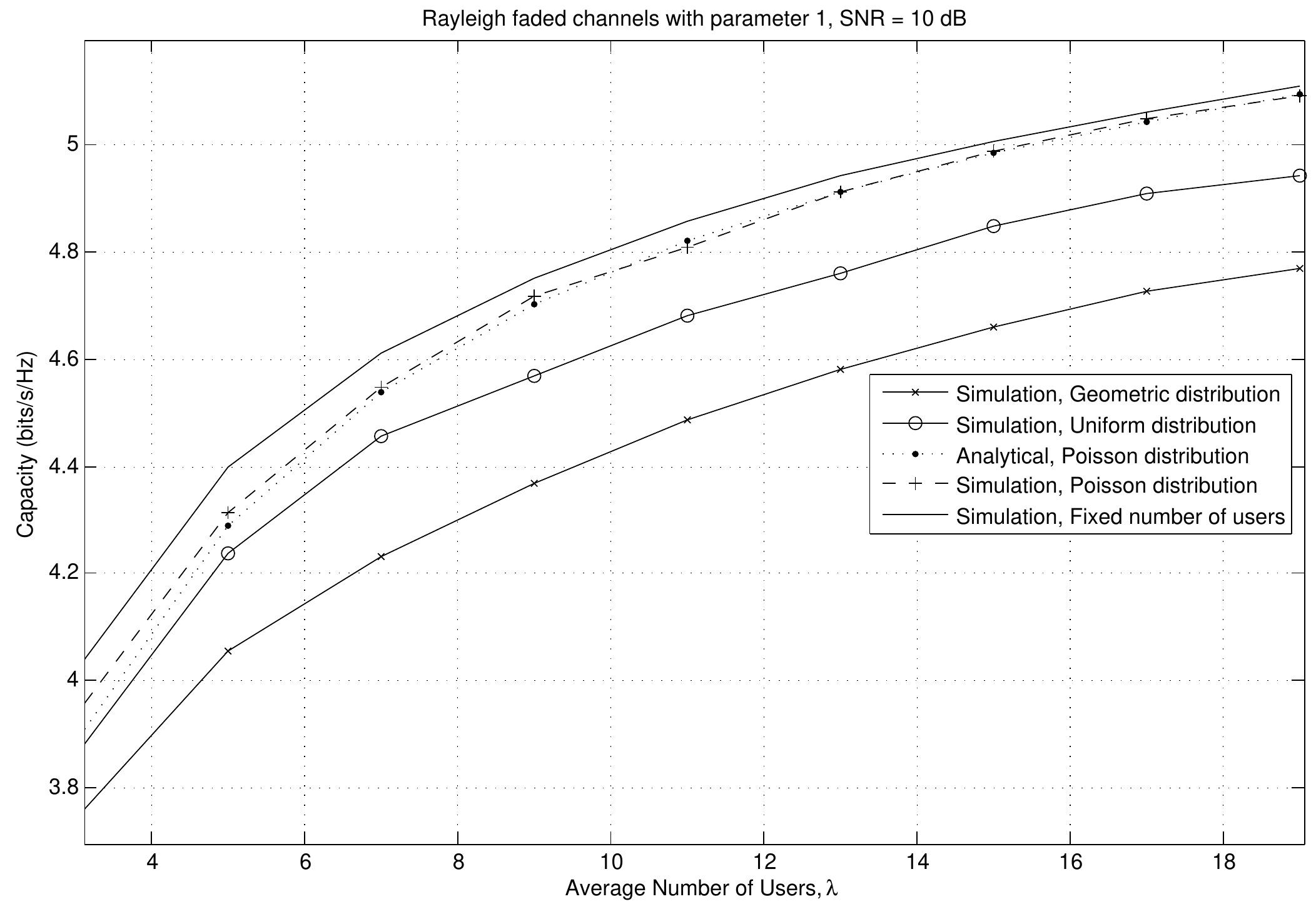}
\caption{Capacity vs. $\lambda$: Rayleigh Fading Channel, average
SNR = 10dB} \label{fig: fig2}
\end{center}
\end{figure}

In Figure \ref{fig: fig2}, the ergodic capacity is plotted against
$\lambda$ for the random cases and $N=\ld$ for the deterministic
case. It is seen that the capacity of the deterministic number of
users system is the highest while for all distributions of
$\mathcal{N}$, the capacity is worse, corroborating our result in
Section \ref{sec: cap_cmd}.

\begin{figure}[!htp]
\begin{center}
\includegraphics[height=9cm,keepaspectratio]{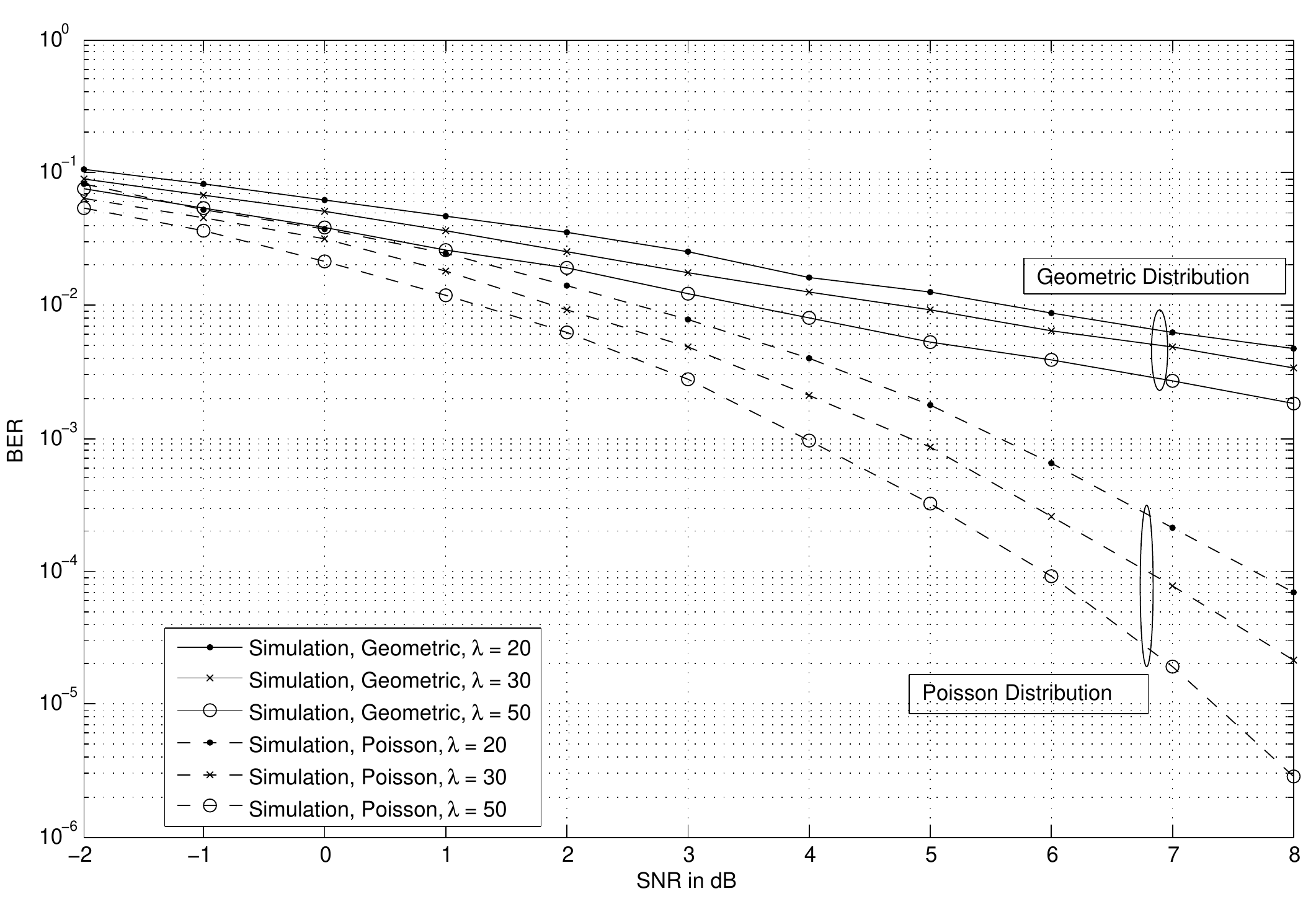}
\caption{Error rate vs. average SNR: Rayleigh Fading Channel}
\label{fig: fig4}
\end{center}
\end{figure}

\begin{figure}[!htp]
\begin{center}
\includegraphics[height=9cm,keepaspectratio]{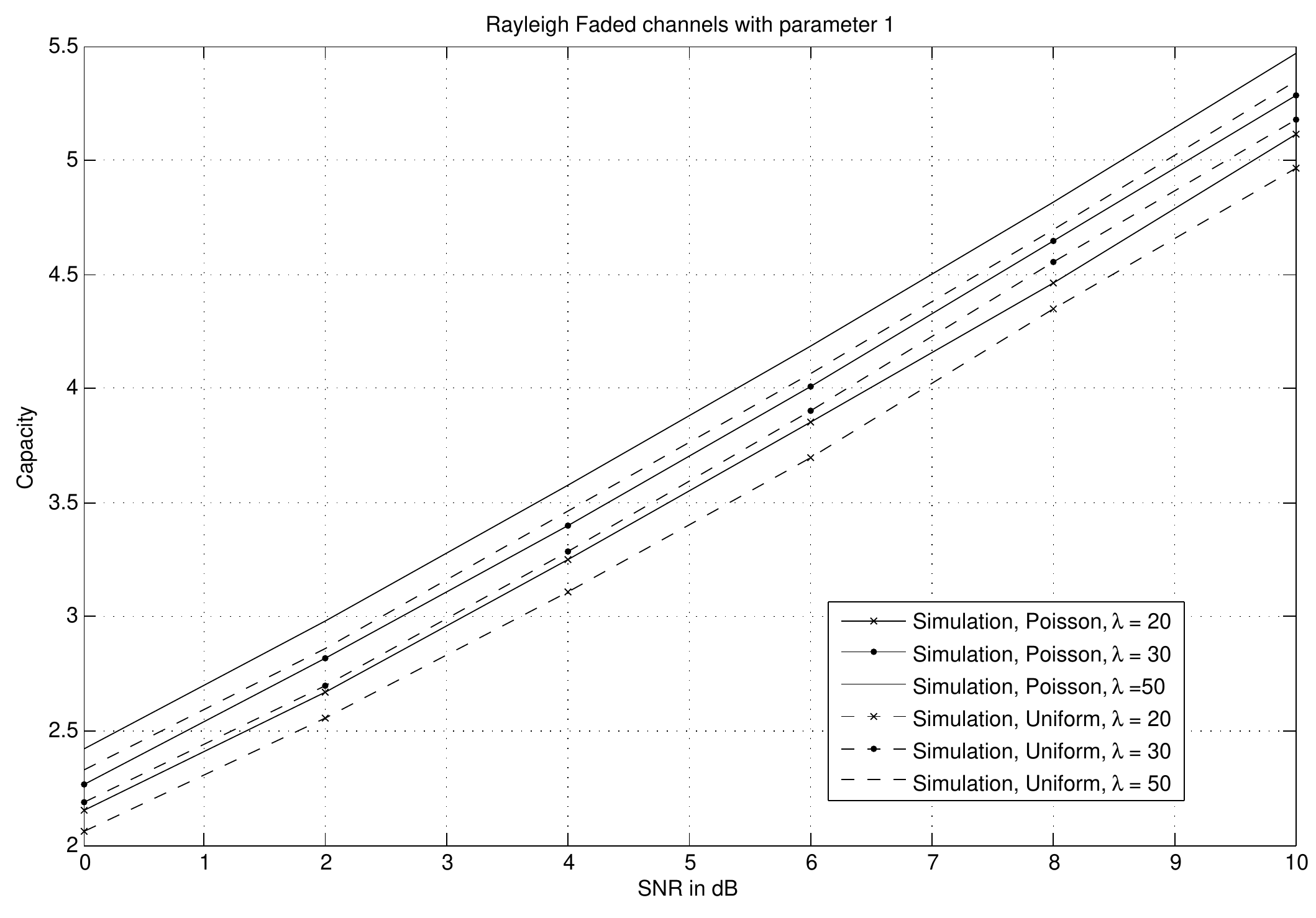}
\caption{Capacity vs. average SNR: Rayleigh Fading Channel}
\label{fig: fig5}
\end{center}
\end{figure}

In Section \ref{sec: LT_ordering}, we showed that Poisson
distributed random variables and geometric distributed random
variables are LT ordered, which also orders their respective average
error rate and ergodic capacities when averaged across the
respective user distributions. In Figures \ref{fig: fig4} and
\ref{fig: fig5} it can be seen that both error rate and capacity
follow their corresponding ordering at all average SNR $\rho$.

\begin{figure}[!htp]
\begin{center}
\includegraphics[height=9cm,keepaspectratio]{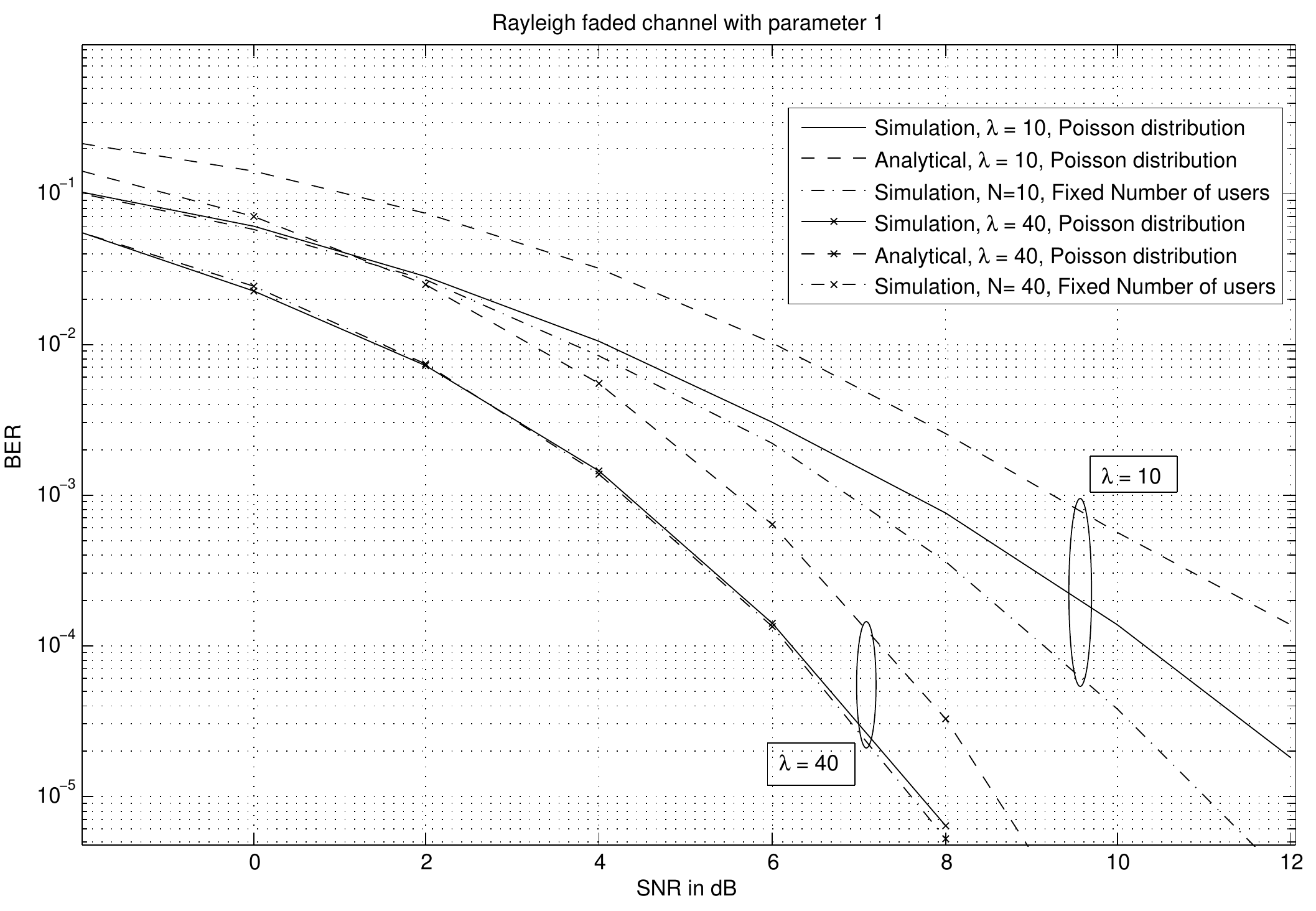}
\caption{Error rate vs. average SNR: Poisson Users and Rayleigh
Fading Channel} \label{fig: fig6}
\end{center}
\end{figure}

In Figure \ref{fig: fig6} the bit error rate versus average SNR for
different average number of users is shown. It can be seen that the
analytical approximation of average error rate with Poisson number
of users derived in \eqref{eqn: avg_ber_poiss_exp2} is within $1$ dB
of the Monte-Carlo simulation result. Following the result in
\eqref{eqn: avg_ber_poiss_exp2} it can also be seen that the larger
the value of $\lambda$, the lower the error rate.

\begin{figure}[!htp]
\begin{center}
\includegraphics[height=9cm,keepaspectratio]{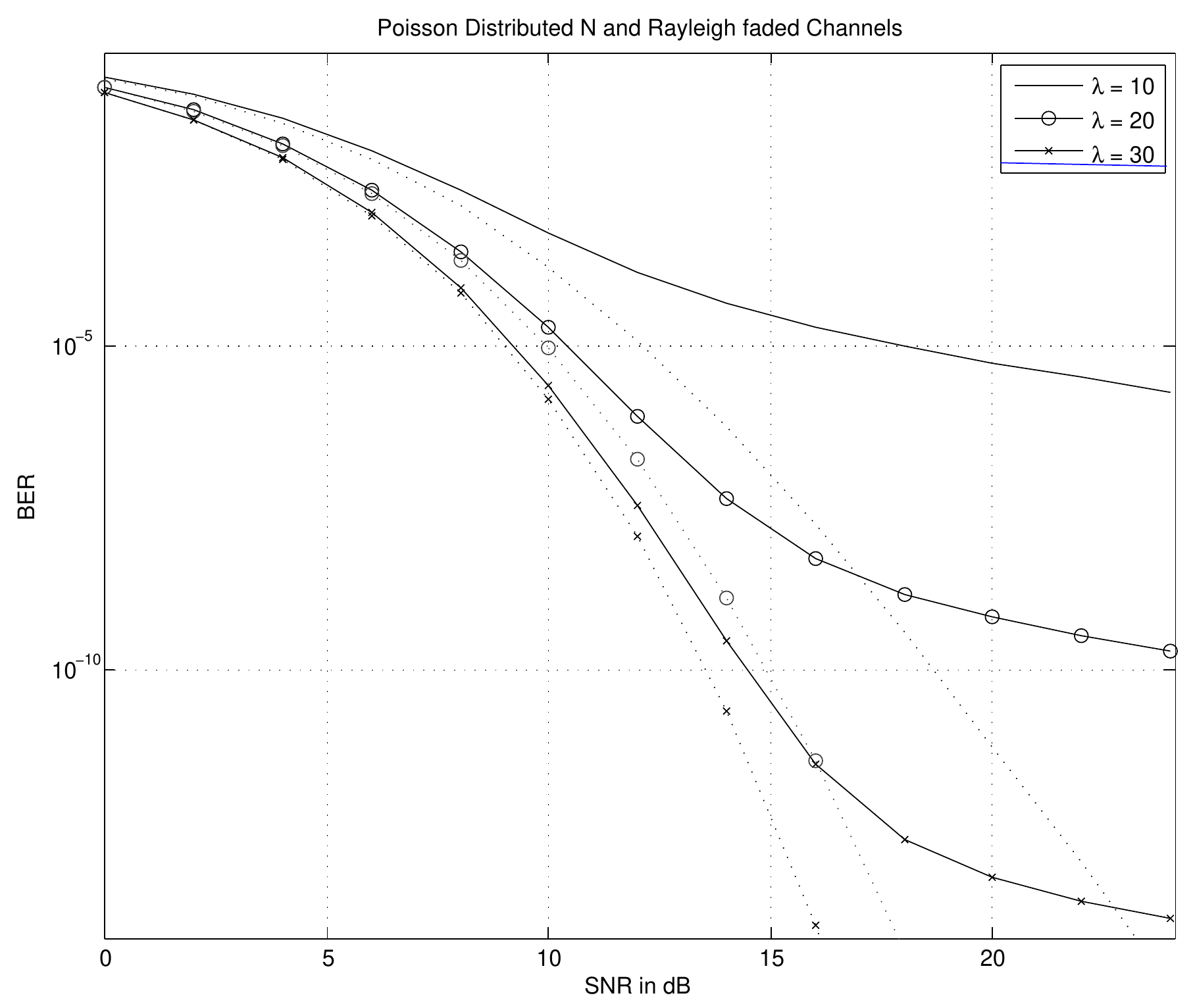}
\caption{Diversity Analysis: Poisson Users and Rayleigh Fading
Channel} \label{fig: fig7}
\end{center}
\end{figure}

In previous simulations we saw that increasing $\lambda$ leads to an
improvement in performance, and for a fixed average SNR the
performance of the system with Poisson distributed number of users
approaches the performance of the system with deterministic number
of users. Figure \ref{fig: fig7} considers a zero-truncated Poisson
distribution and illustrates that at high average SNR, the diversity
order is $k_0 d=1$, verifying Theorem \ref{thm_ser_asympt}. This
leads us to conclude that for low SNR's but sufficiently large
$\lambda$ the performance of the random number of users is nearly
identical to that of the deterministic case. However for high SNR's,
the performance of the random number of users case is significantly
worse due to the loss in diversity order.

\section{Conclusions}\label{sec: conclusions}

Multi-user diversity (MUD) is analyzed for when the number of users
in the system is random. The error rate of MUD systems is proved to
be a completely monotone function of the number of users in the
system, which also implies convexity. Further, ergodic capacity is
shown to have a completely monotone derivative with respect to the
number of users. Using Jensen's inequality, it is shown that the
average error rate and ergodic capacity averaged across fading {\it
and} the number of users will always perform inferior to the
corresponding performance of a system with deterministic number of
users. Further, we provide a method to compare the performance of
the system for different user distributions, using a specific
stochastic ordering based on the Laplace transform of user
distributions.

Importantly, for the MUD system with random number of users, it is
shown that the diversity order is defined by the minimum of the
range of realizations of the number of users. When the number of
users are Poisson distributed, for any user channel fading
distribution, outage probability is shown to converge to the
truncated Gumbel CDF, similar to the case of the deterministic
number of users system. Further, it is proved that the difference
between the error rate performance of the Poisson number of users
system and the deterministic number of users case goes to zero like
$O\left(\overline{\text{P}}_e(\rho, \ld)/\ld \right)$ asymptotically
in the average number of users. As a special case, when the user
fading channels are Rayleigh distributed, a closed-form error rate
expression is provided. Also, the asymptotic scaling law of ergodic
capacity is analyzed and shown to be approximately $\log\left(1+
\rho \log(\ld) \right)$ for large $\ld$. Finally, zero-truncated
Poisson number of users case is shown to not affect our main
conclusions for the common scenario where the number of users is
always positive.

\renewcommand\thesection{Appendix \Alph{section}}
\setcounter{section}{0}

\section{Proof of Theorem \ref{thm_ser_asympt}} \label{thm_ser_asympt_proof}

Since $F_{\gamma}(x)$ is regularly varying, it must be in the form
$F_{\gamma}(x)= x^d l(x)$ where $l(x)$ is slowly varying at $0$. For
a system with $k$ users, the CDF of the channel gain becomes
$F_{\gamma}^k(x)= x^{kd} l^k(x)$. It is easy to verify that $l^k(x)$
is slowly varying at $0$. Therefore, given $\int_0^t
dF_{\gamma}^k(x) \sim t^{kd} l^k(t)$ and $\berx = \al e^{-\eta \rho
x}$, it follows based on Theorem \ref{tbr_th} that
\begin{equation} \label{ser_asympt_exp}
\overline{\text{P}}_e(\rho, k)= \alpha \int_0^{\infty} e^{-\eta \rho
x} dF_{\gamma}^k(x) \sim \alpha \Gamma(k d+ 1) F_{\gamma}^k
(\eta^{-1} \rho^{-1})
\end{equation}
as $\rho \rightarrow \infty$. For the case $\berx = \al Q(\sqrt{\eta
\rho x})$, the asymptotic expression of $\overline{\text{P}}_e(\rho,
k)$ can be derived similarly as follows. Using integration by parts
we obtain
\begin{equation} \label{ser_asympt_qf0}
\overline{\text{P}}_e(\rho, k)= \alpha \int_0^{\infty} Q(\sqrt{\eta
\rho x}) dF_{\gamma}^k(x)= \frac{\alpha \sqrt{\eta \rho}}{2 \sqrt{2
\pi}} \int_0^{\infty} \frac{e^{-\eta \rho x/2}}{\sqrt{x}}
F_{\gamma}^k(x) dx
\end{equation}
Based on \eqref{sv_out_int} we have
\begin{equation}
\int_0^t \frac{1}{\sqrt{x}} F_{\gamma}^k(x) dx \sim l^k(t) \int_0^t
x^{kd-1/2} dx= \frac{1}{kd+1/2} t^{kd+1/2} l^k(t)
\end{equation}
as $t \rightarrow 0$. Then the asymptotic average error rate given
by \eqref{ser_asympt_qf0} becomes
\begin{equation} \label{ser_asympt_qf}
\begin{split}
\overline{\text{P}}_e(\rho, k) &\sim \frac{\alpha \sqrt{\eta
\rho}}{(2kd+1) \sqrt{2 \pi}} \left( \frac{\eta \rho}{2}
\right)^{-kd-1/2} \Gamma (kd+3/2) l^k(2 \eta^{-1} \rho^{-1})\\
&= \frac{\alpha \Gamma (kd+1/2)}{2 \sqrt{\pi}} \left( \frac{\eta
\rho}{2} \right)^{-kd} l^k(2 \eta^{-1} \rho^{-1})= \frac{\alpha
\Gamma (kd+1/2)}{2 \sqrt{\pi}} F_{\gamma}^k (2 \eta^{-1} \rho^{-1})
\end{split}
\end{equation}
as $\rho \rightarrow \infty$, where the asymptotic equality is based
on Theorem \ref{tbr_th} and the second equality is based on
$F_{\gamma}(x)= x^d l(x)$. It can be seen that both
\eqref{ser_asympt_exp} and \eqref{ser_asympt_qf} have the form
$\overline{\text{P}}_e(\rho, k) \sim C_1 F_{\gamma}^k (C_2
\rho^{-1})$ with $C_1$, $C_2$ being constants. Consequently, as
$\rho \rightarrow \infty$, the dominant term in the average error
rate $\E_{\mathcal{N}}\left[ \overline{\text{P}}_e(\rho,
\mathcal{N}) \right] = \sum_{k=k_0}^{\infty}{\pr}
\left[\mathcal{N}=k\right]\overline{\text{P}}_e(\rho, k)$ is the
term with $k=k_0$, which deceases slower than any other term. Using
dominated convergence theorem, we can easily determine $\lim_{\rho
\rightarrow \infty} \E_{\mathcal{N}}\left[
\overline{\text{P}}_e(\rho, \mathcal{N})
\right]/\overline{\text{P}}_e(\rho, k_0)= {\pr}
\left[\mathcal{N}=k_0 \right]$ by exchanging limit and summation.
Consequently, the ratio between the left hand side and the right
hand side of \eqref{ser_asympt} goes to $1$ as $\rho \rightarrow
\infty$, and we thus have the asymptotic average error rate given by
\eqref{ser_asympt}, with $C_1$, $C_2$ given as in the theorem.

\section{Proof of Theorem \ref{cap_scaling}} \label{cap_scaling_proof}

For Poisson distributed $\mathcal{N}$ and Rayleigh faded channels,
using integration by parts and the CDF in \eqref{eqn:
cdf_poisson_exp}, the capacity of the system can be written as,
\begin{equation} \label{eqn: asymp_cap_eq1}
\ecap = \rho \int_0^{\infty} \frac{1-{e^{-\ld e^{-x}}}}{1+ \rho x}
\dr x.
\end{equation}
Defining $y := e^{-x}$ and integrating by substitution,
\begin{align}
\ecap &=  \int_0^1 \frac{1-e^{-\ld y}}{1-\rho \log(y)} \left(\frac{\rho}{y} \right) \dr y  \nonumber \\
&= \int_0^{\sqrt{\log(\ld)}/\ld} \frac{1-e^{-\ld y}}{1-\rho \log(y)}
\left(\frac{\rho}{y} \right) \dr y + \int_{\sqrt{\log(\ld)}/\ld}^1
\frac{\rho (1-e^{-\ld y})}{y(1-\rho \log(y))} \dr y \label{eqn:
cap_pexp1}.
\end{align}
For the first term after the second equality in \eqref{eqn:
cap_pexp1}, we have
\begin{equation} \label{cap_term1_bound}
\begin{split}
0 &< \int_0^{\sqrt{\log(\ld)}/\ld} \frac{1-e^{-\ld y}}{1-\rho
\log(y)} \left(\frac{\rho}{y} \right) \dr y<
\int_0^{\sqrt{\log(\ld)}/\ld} \frac{\ld y}{1+\rho \log(\lambda)-
(\rho/2)\log(\log(\lambda))} \left(\frac{\rho}{y} \right) \dr y\\
&=\frac{\rho \sqrt{\log(\ld)}}{1+\rho \log(\lambda)-
(\rho/2)\log(\log(\lambda))},
\end{split}
\end{equation}
by replacing the numerator of the integrand with its upper bound and
the denominator of the integrand with its lower limit. It can be
seen that the upper bound after the equality in
\eqref{cap_term1_bound} yields $O(1/\sqrt{\log(\ld)})$ and has limit
$0$ as $\ld \rightarrow \infty$, implying that the first term should
have limit $0$. The second term in \eqref{eqn: cap_pexp1} has the
bounds given by,
\begin{equation}
\int_{\sqrt{\log(\ld)}/\ld}^1 \frac{\rho \left(
1-e^{-\sqrt{\log(\ld)}} \right)}{y(1-\rho \log(y))} \dr y <
\int_{\sqrt{\log(\ld)}/\ld}^1 \frac{\rho (1-e^{-\ld y})}{y(1-\rho
\log(y))} \dr y< \int_{\sqrt{\log(\ld)}/\ld}^1 \frac{\rho
(1-e^{-\ld})}{y(1-\rho \log(y))} \dr y
\end{equation}
in which the lower and upper bounds are obtained by bounding the
numerator, and they turn out to be $\left( 1-e^{-\sqrt{\log(\ld)}}
\right) \log(1+ \rho \log(\ld)- (\rho/2)\log(\log(\lambda)))$ and
$(1-e^{-\ld}) \log(1+ \rho \log(\ld)- (\rho/2)\log(\log(\lambda)))$
respectively. Also, it can be verified that $\log(1+ \rho \log(\ld)-
(\rho/2)\log(\log(\lambda)))= \log(1+ \rho \log(\ld))+
O(\log(\log(\lambda))/\log(\ld))$ as $\ld \rightarrow \infty$, and
$\lim_{\ld \rightarrow \infty} e^{-\sqrt{\log(\ld)}} \log(1+ \rho
\log(\ld))= \lim_{\ld \rightarrow \infty} e^{-\ld} \log(1+ \rho
\log(\ld))= 0$. Therefore, for a fixed $\rho$, and as $\ld
\rightarrow \infty$ we can express \eqref{eqn: cap_pexp1} as
\eqref{cap1_asympt} considering the fact that
$\log(\log(\lambda))/\log(\ld)$ decays faster than
$1/\sqrt{\log(\ld)}$, completing the proof.

\bibliographystyle{IEEEtran}
\bibliography{adarsh_ref}

\end{document}